\documentclass[conference]{IEEEtran}

\usepackage{multirow}
\usepackage{mathtools}
\usepackage{amsfonts} 
\usepackage{amsmath,bm}

\usepackage{cite}
\usepackage{array}
\usepackage{setspace}
\usepackage{longtable}
\usepackage{threeparttable}

\ifCLASSINFOpdf
\usepackage[pdftex]{graphicx}
\else
\fi

\usepackage[ruled,linesnumbered]{algorithm2e}

\usepackage{makecell}

\usepackage{algpseudocode}

\usepackage{amsmath,amsfonts,amssymb,mathtools}
\usepackage{mathrsfs}
\usepackage{color}
\usepackage[dvipsnames]{xcolor}
\usepackage{amsthm}
\usepackage[normalem]{ulem}
\usepackage{nomencl}
\usepackage{etoolbox}
\usepackage[nocomma]{optidef}
\theoremstyle{definition}

\makenomenclature

\graphicspath{figure}

\begin{document}

\title{
Frequency-Security-Aware Real-Time Economic Dispatch with Data Center UPS Flexibility for Inertia and Primary Frequency Response Services
}

\author{\IEEEauthorblockN{Shengyi Wang}
\IEEEauthorblockA{\textit{Department of Electrical \& Computer Engineering}\\ 
University of Arkansas at Little Rock\\
Little Rock, AR USA\\
Email:
swang@ualr.edu}
\and
\IEEEauthorblockN{Liang Du}
\IEEEauthorblockA{
\textit{Department of Electrical \& Computer Engineering}\\
Villanova University\\
Villanova, PA USA\\
Email: liang.du@villanova.edu}}


\maketitle
\vspace{-1.2em}

\begin{abstract}
The increasing penetration of renewable energy and the rapid growth of AI data centers are introducing significant challenges to power system frequency security.
There is an urgent need to identify grid-compatible resources capable of providing fast frequency support.
This paper proposes a frequency-security-aware real-time economic dispatch (RTED) model that enables data center (DC) uninterruptible power supply (UPS) systems to provide steady-state power balancing, inertia response, and primary frequency response (PFR) services. Specifically, tractable analytical frequency-security constraints are derived and incorporated into the RTED framework through second-order cone programming (SOCP) formulation. The proposed model jointly optimizes energy, reserve, and inertia to minimize system operation cost while preventing under-frequency load shedding (UFLS). Numerical validation on a modified IEEE 39-bus system demonstrates that the proposed RTED model with UPS flexibility can economically support system operation and significantly improve frequency security under active-power deficit disturbances.

\end{abstract}

\begin{IEEEkeywords}
Data center UPS systems, Power flexibility, 
Frequency-constrained
economic operation,
Frequency support.
\end{IEEEkeywords}

\IEEEpeerreviewmaketitle

\section{Introduction}
The rapid transformation of modern power systems is reshaping both the supply and demand sides of the electricity grid. On the supply side, the ongoing low-carbon transition has accelerated the large-scale integration of inverter-based renewable energy resources, which are gradually replacing conventional synchronous generators (SGs). Although this transition significantly reduces carbon emissions, it also decreases the physical inertia and PFR capability inherently provided by synchronous machines. On the demand side, the explosive growth of AI, big data analytics, and cloud computing is driving the rapid expansion of large-scale DCs\cite{chen2025electricity}, making them increasingly important electricity consumers. The simultaneous evolution of generation and load characteristics introduces more complicated power system frequency dynamics and increases the challenges of maintaining frequency security under uncertainty\cite{yan2023frequency}.

To enhance the frequency security of modern power systems, additional frequency support resources are required\cite{he2024analysis}. However, investing in dedicated frequency support assets often leads to considerable capital and operating costs, which may not be economically attractive. Consequently, increasing attention has been directed toward exploiting the inherent flexibility of existing grid-connected resources to provide ancillary services \cite{wang2025bayesian}. In recent years, several studies have highlighted the grid compatibility of DCs and recognized them as promising flexible resources for power systems\cite{ratnayake2025grid,chen2025spatial,jin2025unlocking}. By modulating computational workloads and electricity demand, DCs can interact with the power grid and provide various ancillary services, including frequency regulation\cite{zhang2021hpc}. 

Among the various components inside a DC, the UPS system \cite{tao2025fast} is particularly attractive for frequency support. A DC UPS system is a battery-based backup power system designed to ensure the uninterrupted operation of critical IT and cooling loads during grid disturbances or outages. Equipped with advanced power-electronic control strategies \cite{padmanabhan2022procurement}, such as grid-forming control \cite{hu2022grid}, the energy stored in UPS batteries can be flexibly charged or discharged to provide bidirectional frequency response while emulating virtual inertia. Compared with conventional synchronous generators, UPS systems typically exhibit shorter response delays and faster ramping characteristics, making them well suited for providing inertia and PFR services during the early stage of post-disturbance frequency evolution. Moreover, enhancing the overall frequency stability of the power system can also improve the frequency ride-through capability and operational resilience of DCs.

Existing studies have investigated the use of DC UPS flexibility for frequency support. For example, \cite{al2021potential} proposes a model predictive control-based method to coordinate UPS systems and workload for minimizing the frequency nadir while reducing control effort, and \cite{kaur2019adaptive} designs an adaptive hierarchical control framework to coordinate UPS systems and workload for secondary frequency regulation. Nevertheless, several important research gaps remain. 
Existing studies mainly focus on control design for coordinating UPS systems and workloads, whereas 
limited attention has been paid to
integrating UPS-based fast frequency support into power system operation frameworks
\cite{she2023virtual} such as unit commitment, RTED, and automatic generation control. Such integration is crucial for economically dispatching UPS resources to support frequency security while ensuring coordinated operation with conventional SGs.

To bridge the above gaps, this paper proposes a frequency-security-aware RTED model that enables DC UPS systems to provide steady-state power balancing, inertia and PFR services. 
The main contributions of this work are twofold. First, tractable analytical frequency-security constraints are derived by explicitly modeling the frequency-support capability of DC UPS systems, and a participation qualification condition is established to identify UPS systems suitable for ancillary service provision. Second, a novel operation model is developed to jointly optimize power, reserve, and inertia for system operation cost minimization with the enhanced frequency security under active-power deficit disturbances. 


\section{Frequency-Security Constraints Modeling}
Let $t \in \mathcal{T} := \{0,1,\dots,T-1\}$ denote the scheduling interval index, and let $\tau$ denote the transient-time variable.  
Since this work only considers under-frequency events, the disturbance term satisfies $\Delta p_t^{\text{DIS}} > 0, \forall t \in \mathcal{T}$
where $\Delta p_t^{\text{DIS}}$ denotes the equivalent active-power deficit introduced by the disturbance event.

Let $\mathcal{N}_g$ and $\mathcal{N}_{dc}$ denote the index sets of synchronous generators and DC UPS systems, respectively. With appropriate power-electronic control strategies, a DC UPS system can emulate virtual inertia $M_{u_j,t}$ and contribute to frequency stability.
Then the system inertia can be defined as follows
\begin{equation}
\begin{aligned}
M_{\text{sys},t}
&= \sum\limits_{i \in \mathcal{N}_g} M_{g_i} \frac{S_{g_i}}{S_b}
+ \sum\limits_{j \in \mathcal{N}_{dc}} M_{u_j,t} \frac{S_{u_j}}{S_b}
\end{aligned}
\label{eq:system_inertia}
\end{equation}
where $M_{g_i}$ is the inertia constant of synchronous generator $i$; $S_{g_i}$ and $S_{u_j}$ are the rated apparent power for synchronous generator $i$ and UPS $j$, respectively; and $S_b$ is the selected system power base.
Since this work considers a one-hour scheduling horizon with predetermined unit commitment decisions, the inertia contribution from each SG remains unchanged during the scheduling horizon. Therefore, $M_{g_i}$ is modeled as a constant parameter for all $t\in\mathcal{T}$.

Let $\hat{D}_{l,t}$
denote the equivalent load damping coefficient at scheduling interval $t$.
Following a disturbance $\Delta p_t^{\text{DIS}}$, the system frequency dynamics can be characterized by a first-order linear differential equation with constant coefficients
\begin{equation}
\begin{aligned}
&M_{\text{sys},t} \frac{d \Delta f_t(\tau)}{d \tau} + 
\hat{D}_{l,t} \Delta f_t(\tau) = \\
&\sum\limits_{i \in \mathcal{N}_g} \Delta p_{g_i,\tau}^{\text{PFR}} 
+
\sum\limits_{j \in \mathcal{N}_{dc}} \Delta p_{u_j,\tau}^{\text{PFR}}
-
\Delta p_t^{\text{DIS}}, \quad \forall{t \in \mathcal{T}}
\end{aligned}
\label{eq:frequency_dynamic_response}
\end{equation}
where $\Delta f_t(\tau)$ denotes the system frequency deviation at transient time $\tau$ at scheduling interval $t$;
and $\Delta p_{g_i,\tau}^{\text{PFR}}$ and $\Delta p_{u_j,\tau}^{\text{PFR}}$ denote the PFR power delivered by SG $i$ and UPS $j$, respectively, which can be modeled as a linear ramp trajectory 
\begin{equation}
\begin{aligned}
\Delta p_{g_i,\tau}^{\text{PFR}} =  \left\{ \begin{array}{cl}
0 & : \ \tau \leq \Phi_{db}^{G} \\
R_{g_i,t} \frac{\tau - \Phi_{db}^{G}}{\Phi^G} & : \ \Phi_{db}^{G} \leq \tau \leq \Phi_{db}^{G}+\Phi^{G} \\
R_{g_i,t} & : \ \tau \geq \Phi_{db}^{G}+\Phi^{G}
\end{array} \right.
\end{aligned}
\label{eq:ramp_model_generator}
\end{equation}
\begin{equation}
\begin{aligned}
\Delta p_{u_j,\tau}^{\text{PFR}} =  \left\{ \begin{array}{cl}
0 & : \ \tau \leq \Phi_{db}^{U} \\
R_{u_j,t} \frac{\tau - \Phi_{db}^{U}}{\Phi^{U}} & : \ \Phi_{db}^{U} \leq \tau \leq \Phi_{db}^{U}+\Phi^{U} \\
R_{u_j,t} & : \ \tau \geq \Phi_{db}^{U}+\Phi^{U}
\end{array} \right.
\end{aligned}
\label{eq:ramp_model_ups}
\end{equation}
where $R_{g_i,t}$ and $R_{u_j,t}$ denote the scheduled PFR reserve capacities of synchronous generator $i$ and UPS $j$ at scheduling interval $t$, respectively; $\Phi_{db}^{G}$ and $\Phi_{db}^{U}$ represent the response deadband delays of synchronous generators and UPSs, respectively; and $\Phi^{G}$ and $\Phi^{U}$ denote the ramp-up durations required for synchronous generators and  UPSs to reach their scheduled PFR reserve capacities. 

\begin{figure}[ht]
\centering
\includegraphics[width=0.38\textwidth]{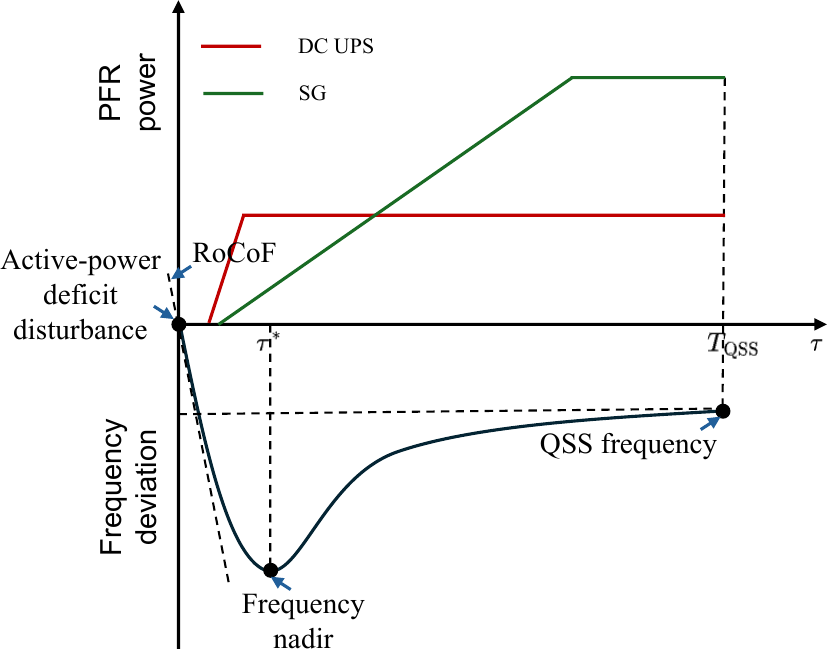}
\caption{Illustration of the post-disturbance frequency evolution and the assumed PFR trajectories of DC UPS systems and SGs.}
\label{fig:frequency_response_process}
\end{figure}

As shown in Fig.~\ref{fig:frequency_response_process}, both DC UPS systems and SGs contribute to the PFR following an active-power deficit disturbance. Owing to their power-electronics-interfaced architecture, DC UPS systems typically exhibit shorter response deadband delays and faster ramp-up rates than conventional SGs, i.e., $\Phi_{db}^{U}\leq\Phi_{db}^{G}$ and $\Phi^{U}\leq\Phi^{G}$. Consequently, UPS systems can rapidly inject active power during the early stage of the post-disturbance frequency evolution, while SGs gradually increase their PFR outputs through governor actions. After reaching their scheduled PFR reserve capacities, both resources are assumed to maintain their outputs until the system reaches the quasi-steady-state (QSS) frequency.

To ensure secure and stable power system operation following a disturbance, the system operator must maintain the post-fault frequency indicators within prescribed safety thresholds. As illustrated in Fig.~\ref{fig:frequency_response_process}, three frequency-security indicators are of particular interest, namely the rate of change of frequency (RoCoF), the frequency nadir, and the QSS frequency. The RoCoF characterizes the rate at which the system frequency changes with respect to time following a disturbance. The frequency nadir represents the minimum system frequency reached before the PFR resources arrest the frequency decline, while the QSS frequency characterizes the final frequency equilibrium established after the completion of the PFR process and before the secondary frequency control becomes dominant.

Based on the assumed post-disturbance frequency response process, tractable analytical expressions for the RoCoF, frequency nadir, and QSS frequency constraints are derived in the following subsections.

\subsection{RoCoF Constraints}
At the onset of an active-power deficit disturbance, the PFRs from DC UPS systems and SGs have not yet been effectively deployed because of their finite response delays. Moreover, since the system frequency deviation is initially zero, the load damping effect can be neglected. Therefore, the initial frequency dynamics are governed solely by the equivalent system inertia, and the maximum RoCoF occurs immediately after the disturbance. Consequently, the initial RoCoF, which represents the maximum RoCoF during the post-disturbance frequency evolution, is adopted for frequency-security assessment and is constrained as
\begin{equation}
\begin{aligned}
\mathrm{RoCoF}_{\max,t}
&= \bigg|\left.\frac{d \Delta f_t(\tau)}{d \tau}\right|_{\tau \to 0^+}\bigg| f_N \\
&= \frac{|\Delta p_t^{\text{DIS}}| f_N}{M_{\text{sys},t}}
= \frac{\Delta p_t^{\text{DIS}} f_N}{M_{\text{sys},t}}
\leq \mathrm{RoCoF}_{\lim}
\end{aligned}
\label{eq:max_RoCoF}
\end{equation}
where $f_N$ is the nominal system frequency and $\mathrm{RoCoF}_{\lim}$ is the maximum allowable RoCoF. 

\subsection{Frequency Nadir Constraints}
To obtain a tractable analytical expression for the frequency nadir, the load damping term is neglected during the transient nadir derivation. 
Let $\tau^{*}$ denote the instant at which the frequency nadir is reached.
Since the frequency deviation satisfies
$\Delta f(0)=0$, integrating the frequency dynamic equation
(\ref{eq:frequency_dynamic_response}) from $\tau=0$ to
$\tau=\tau^{*}$ yields
\begin{equation}
\begin{aligned}
\Delta f(\tau^{\ast})
&=
\frac{1}{M_{\text{sys},t}}\bigg\{
\sum\limits_{i \in \mathcal{N}_g}
\int_{0}^{\tau^{\ast}} \Delta p_{g_i,\tau}^{\text{PFR}} d\tau \\
&\quad +
\sum\limits_{j \in \mathcal{N}_{dc}}
\int_{0}^{\tau^{\ast}} \Delta p_{u_j,\tau}^{\text{PFR}} d\tau
-
\int_{0}^{\tau^{\ast}} \Delta p_t^{\text{DIS}} d\tau
\bigg\}
\end{aligned}
\label{eq:nadir2}
\end{equation}

As shown in Fig.~\ref{fig:frequency_response_process},
DC UPS systems typically complete their PFR ramping process before the frequency nadir is reached, whereas SGs are still in the ramping stage.
Therefore, this work assumes
$\Phi_{db}^{G} \leq \tau^{\ast} \leq \Phi_{db}^{G} + \Phi^{G}$ and 
$\tau^{\ast} \geq \Phi_{db}^{U} + \Phi^{U}$
which indicates that the UPS PFR has reached its scheduled reserve capacity while the SGs continue ramping toward their scheduled PFR reserve capacities.
Under this assumption, the analytical form of
Eq.~\eqref{eq:nadir2} can be expressed as
\begin{equation}
\begin{aligned}
\Delta f(\tau^{\ast})
&= \frac{1}{M_{\text{sys},t}}\bigg\{
\frac{\left(\tau^{\ast} - \Phi_{db}^{G}\right)^2}{2\Phi^G} \sum\limits_{i \in \mathcal{N}_g} R_{g_i,t} \\
&\quad + \left(\tau^{\ast} - \Phi_{db}^{U} - \frac{\Phi^U}{2}\right)
\sum\limits_{j \in \mathcal{N}_{dc}}
R_{u_j,t}
-
\Delta p_t^{\text{DIS}} \tau^{\ast}
\bigg\}
\end{aligned}
\label{eq:nadir3}
\end{equation}

At $\tau=\tau^{\ast}$, the following approximation is adopted
\begin{equation}
\begin{aligned}
\sum\limits_{i \in \mathcal{N}_g} \Delta p_{g_i,\tau^{\ast}}^{\text{PFR}} 
+
\sum\limits_{j \in \mathcal{N}_{dc}} \Delta p_{u_j,\tau^{\ast}}^{\text{PFR}}
=
\Delta p_t^{\text{DIS}}
\end{aligned}
\label{eq:nadir4}
\end{equation} 
Substituting the ramp-based PFR models in Eqs.~(\ref{eq:ramp_model_generator}) and (\ref{eq:ramp_model_ups}) into Eq.~(\ref{eq:nadir4}), the nadir occurrence time can be obtained as
\begin{equation}
\begin{aligned}
\tau^{\ast} = \Phi_{db}^{G} + \frac{(\Delta p_t^{\text{DIS}}-\sum\limits_{j \in \mathcal{N}_{dc}}
R_{u_j,t}) \Phi^{G}}{\sum\limits_{i \in \mathcal{N}_g} R_{g_i,t}}
\end{aligned}
\label{eq:nadir5}
\end{equation}

For notational simplicity, the aggregate PFR reserve capacities of SGs and DC UPS systems are defined as
\begin{equation}
\begin{aligned}
R_{G,t}
= \sum\limits_{i \in \mathcal{N}_g} R_{g_i,t}, \quad
R_{U,t}
= \sum\limits_{j \in \mathcal{N}_{dc}} R_{u_j,t}
\end{aligned}
\label{eq:aggregate_pfr_capacity}
\end{equation}
Substituting Eq.~\eqref{eq:nadir5} into Eq.~\eqref{eq:nadir3} and using the aggregate
PFR definitions in Eq.~(\ref{eq:aggregate_pfr_capacity}), the frequency nadir can be analytically expressed as
\begin{equation}
\begin{aligned}
\Delta f(\tau^{\ast})
&= \frac{1}{M_{\text{sys},t}}\bigg\{
R_{U,t}\left(\Phi_{db}^{G} - \Phi_{db}^{U} - \frac{\Phi^U}{2}\right)
- \Delta p_t^{\text{DIS}} \Phi_{db}^{G} \\
&\quad - \frac{\Phi^G}{2R_{G,t}}
\left(\Delta p_t^{\text{DIS}} - R_{U,t}\right)^2
\bigg\}
\end{aligned}
\label{eq:nadir6}
\end{equation}

To enable DC UPS systems to participate in inertia and PFR services through the proposed ED model, this work first establishes a participation qualification condition, i.e., $\Phi_{db}^{G} \leq \Phi_{db}^{U}+\frac{\Phi^U}{2}$.
The term $\Phi_{db}^{U}+\frac{\Phi^U}{2}$ represents the effective average response time of the DC UPS systems, corresponding to the midpoint of the linear ramping process in Eq.~(\ref{eq:ramp_model_ups}). Therefore, the above condition implies that the DC UPS systems can provide effective frequency support before or around the time when conventional generators begin their PFR actions. 
Meanwhile, this condition can be used to identify qualified DC UPS systems for participation in the existing ancillary service market.
In practice, such a condition is achievable by properly designing the UPS converter control strategy. Under the active-power deficit disturbance considered in this work, where $\Delta p_t^{\text{DIS}}\geq 0$, this condition guarantees that all terms in Eq.~(\ref{eq:nadir6}) is non-positive, thereby ensuring
$\Delta f(\tau^*) \leq 0$,
which confirms that Eq.~(\ref{eq:nadir6}) indeed corresponds to an approximation of the system frequency nadir.

To prevent the activation of under-frequency load shedding (UFLS), a frequency nadir constraint is imposed as
\begin{equation}
\begin{aligned}
|\Delta f(\tau^{\ast})| f_{N} &= \frac{f_N}{M_{\text{sys},t}} 
\bigg\{
\frac{\Phi^G}{2R_{G,t}}
\left(\Delta p_t^{\text{DIS}} - R_{U,t}\right)^2
+ \Delta p_t^{\text{DIS}} \Phi_{db}^{G} \\
&- R_{U,t}\left(\Phi_{db}^{G} - \Phi_{db}^{U} - \frac{\Phi^U}{2}\right)
\bigg\} \leq \Delta f_{\lim}
\end{aligned}
\label{eq:nadir7}
\end{equation}
where $\Delta f_{\lim}$ is the maximum allowable frequency deviation magnitude. 

Eq.~(\ref{eq:nadir7}) can be rewritten as 
\begin{equation}
\begin{aligned}
\left(
\Delta p_t^{\text{DIS}}
-
R_{U,t}
\right)^2
\le
R_{G,t}z_t
\end{aligned}
\label{eq:nadir8}
\end{equation}
where
\begin{equation}
\left\{
\begin{aligned}
z_t &= \frac{2}{\Phi^{G}}
\biggl[
\frac{\Delta f_{\lim}}{f_N} M_{\text{sys},t}
-
\Delta p_t^{\text{DIS}}\Phi_{db}^{G} \\
&\quad +
R_{U,t}
\left(
\Phi_{db}^{G}
-
\Phi_{db}^{U}
-
\frac{\Phi^{U}}{2}
\right)
\biggr], \\
z_t &\geq 0 .
\end{aligned}
\right.
\label{eq:nadir9}
\end{equation}

Eq.~\eqref{eq:nadir8} provides an alternative interpretation of the frequency nadir requirement. The quantity $(\Delta p_t^{\text{DIS}}-R_{U,t})$ represents the residual power deficit that remains uncovered by the UPS PFR. A larger residual deficit requires either a larger reserve capacity from the SG PFR $R_{G,t}$ or a larger effective system inertia represented by $z_t$. Therefore, Eq.~\eqref{eq:nadir8} explicitly characterizes the trade-off among UPS PFR, SG PFR, and system inertia in maintaining frequency security.

Moreover, under the above-mentioned participation qualification condition, the auxiliary variable $z_t$ becomes a meaningful physical constraint rather than a purely mathematical requirement. Since $z_t$ decreases as the procured UPS PFR reserve capacity increases, the condition $z_t \ge 0$ imposes an implicit upper bound on $R_{U,t}$. This indicates that the procured UPS PFR reserve capacity must be coordinated with the available system inertia and disturbance magnitude to ensure that the frequency nadir requirement remains feasible.

To ensure model tractability, Eq.~(\ref{eq:nadir8}) is reformulated as an equivalent SOC constraint 
\begin{equation}
\begin{aligned}
\left\|
\begin{bmatrix}
2\left(
\Delta p_t^{\text{DIS}}
-
R_{U,t}
\right)
\\
R_{G,t}-z_t
\end{bmatrix}
\right\|_2
\le
R_{G,t}+z_t.
\end{aligned}
\label{eq:nadir10}
\end{equation}

\subsection{Quasi-steady-state Frequency Constraints}
To ensure that sufficient PFR capacity is scheduled to cover the considered active-power deficit by the QSS stage, the aggregate PFR capacity should satisfy
\begin{equation}
\begin{aligned}
R_{G,t} + R_{U,t} \geq \Delta p_t^{\text{DIS}}  
\end{aligned}
\label{eq:nadir11}
\end{equation}

\section{Proposed Economic Dispatch Model}
\begin{figure}[ht]
\centering
\includegraphics[width=0.47\textwidth]{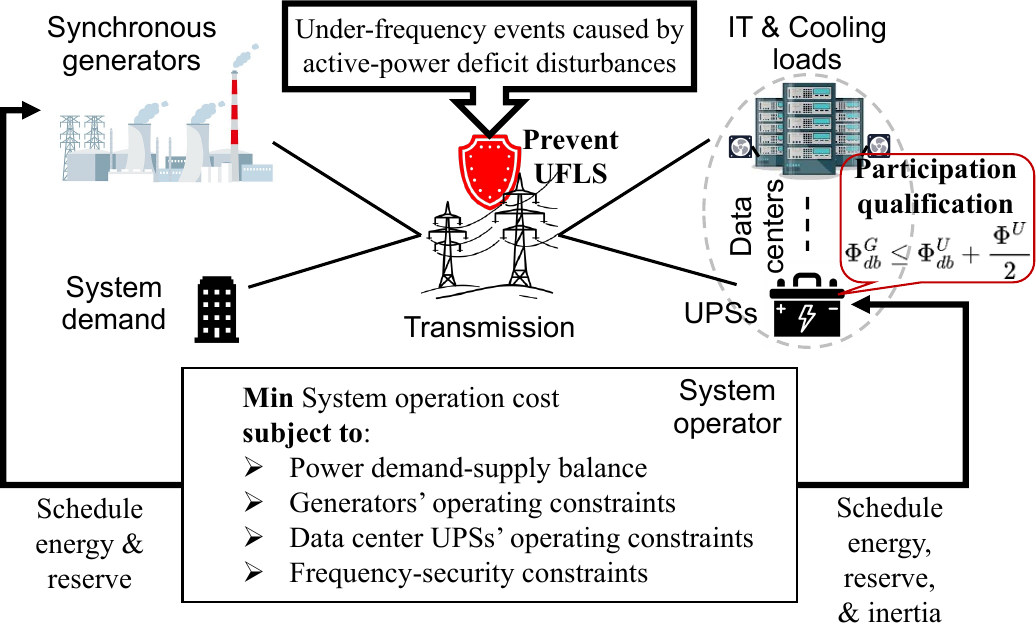}
\caption{Illustration of the proposed frequency-security-aware RTED model.}
\label{fig:model}
\end{figure}

As shown in Fig.~\ref{fig:model}, assuming that DC UPS systems satisfy the above-mentioned participation qualification conditions, an hour-ahead frequency-security-aware RTED model is proposed to jointly optimize the operation of SGs and DC UPS systems based on the hour-ahead net-load forecast $\hat{p}_{n,t}^{\text{HA}}$ over the scheduling horizon, thereby enabling economically efficient system operation while mitigating the risk of UFLS caused by active-power deficit disturbances.
Let $p_{g_i,t}$, $p_{u_j,t}$, and $e_{j,t}$ denote the scheduled power output of SG $i$, the scheduled power output of UPS $j$, and the state of charge (SoC) of UPS $j$, respectively.
The detailed mathematical formulation is presented as follows.
\begin{equation}
\begin{aligned}
&\min\limits_{\Xi_g,\, \Xi_{dc},\Xi_{agg}} 
\sum\limits_{t \in \mathcal{T}} 
[\sum\limits_{i \in \mathcal{N}_g}
\left(a_{g_i} p_{g_i,t} + b_{g_i} R_{g_i,t}\right) \\
&+ \sum\limits_{j \in \mathcal{N}_{dc}}
\left(a_{u_j} p_{u_j,t}^2 + b_{u_j} R_{u_j,t} + c_{u_j} M_{u_j,t}\right)] \Delta t \; \quad \mathrm{s.t.}
\end{aligned}
\label{eq:ed_model1}
\end{equation}
\begin{equation}
\begin{aligned}
\sum\limits_{i \in \mathcal{N}_g} p_{g_i,t} + 
\sum\limits_{j \in \mathcal{N}_{dc}} p_{u_j,t} = \hat{p}_{n,t}^{\text{HA}}, \quad \forall{t \in \mathcal{T}}
\end{aligned}
\label{eq:ed_model2}
\end{equation}
\begin{equation}
\begin{aligned}
p_{g_i,t} + R_{g_i,t} \leq \overline{p}_{g_i}, \quad \forall{i \in \mathcal{N}_g}, \forall{t \in \mathcal{T}}
\end{aligned}
\label{eq:ed_model3}
\end{equation}
\begin{equation}
\begin{aligned}
p_{g_i,t} \geq \underline{p}_{g_i}, \quad \forall{i \in \mathcal{N}_g}, \forall{t \in \mathcal{T}}
\end{aligned}
\label{eq:ed_model4}
\end{equation}
\begin{equation}
\begin{aligned}
0 \leq R_{g_i,t} \leq \overline{R}_{g_i}, \quad \forall{i \in \mathcal{N}_g}, \forall{t \in \mathcal{T}}
\end{aligned}
\label{eq:ed_model5}
\end{equation}
\begin{equation}
\begin{aligned}
R_{G,t} \geq \underline{R}_{G,t}, \quad \forall{t \in \mathcal{T}} 
\end{aligned}
\label{eq:ed_model6}
\end{equation}
\begin{equation}
\begin{aligned}
&\left\{ \begin{array}{cl}
(\ref{eq:system_inertia}),
(\ref{eq:max_RoCoF}), (\ref{eq:aggregate_pfr_capacity}), (\ref{eq:nadir9}) - (\ref{eq:nadir11}) \\
R_{U,t} \leq \Delta p_t^{\text{DIS}} \\
(\Delta p_t^{\text{DIS}}-R_{U,t})\Phi^{G}
\geq
(\Phi_{db}^{U}+\Phi^{U}-\Phi_{db}^{G})R_{G,t}
\end{array} \right., \\
&\forall t\in\mathcal{T}\ \text{such that}\ \Delta p_t^{\text{DIS}}>0.
\end{aligned}
\label{eq:ed_model7}
\end{equation}
\begin{equation}
\begin{aligned}
-\underline{p}_{u_j} \leq p_{u_j,t} \leq \overline{p}_{u_j}, 
\quad \forall{j \in \mathcal{N}_{dc}}, \forall{t \in \mathcal{T}}
\end{aligned}
\label{eq:ed_model8}
\end{equation}
\begin{equation}
\begin{aligned}
p_{u_j,t} + R_{u_j,t}
+ \frac{M_{u_j,t} S_{u_j} \cdot \mathrm{RoCoF}_{\lim}}{S_b f_N}
&\leq \overline{p}_{u_j}, \\
&\forall j \in \mathcal{N}_{dc},\; \forall t \in \mathcal{T}
\end{aligned}
\label{eq:ed_model9}
\end{equation}
\begin{equation}
\begin{aligned}
R_{u_j,t} \geq 0, 
\forall{j \in \mathcal{N}_{dc}}, \forall{t \in \mathcal{T}}
\end{aligned}
\label{eq:ed_model10}
\end{equation}
\begin{equation}
\begin{aligned}
0 \leq M_{u_j,t} \leq \overline{M}_{u_j}, 
\forall{j \in \mathcal{N}_{dc}}, \forall{t \in \mathcal{T}}
\end{aligned}
\label{eq:ed_model11}
\end{equation}
\begin{equation}
\begin{aligned}
e_{j,t+1} = e_{j,t} - \frac{p_{u_j,t} \Delta t}{E_j}, \quad
\forall{j \in \mathcal{N}_{dc}}, \forall{t \in \mathcal{T}}
\end{aligned}
\label{eq:ed_model12}
\end{equation}
\begin{equation}
\begin{aligned}
\underline{e}_j \leq e_{j,t+1} \leq \overline{e}_j, \quad
\forall{j \in \mathcal{N}_{dc}}, \forall{t \in \mathcal{T}}
\end{aligned}
\label{eq:ed_model13}
\end{equation}
\begin{equation}
\begin{aligned}
e_{j,T} = e_{j,0}, \quad
\forall{j \in \mathcal{N}_{dc}}
\end{aligned}
\label{eq:ed_model14}
\end{equation}
\begin{equation}
\begin{aligned}
&e_{j,t+1} 
\geq \frac{M_{u_j,t} S_{u_j} \cdot \mathrm{RoCoF}_{\lim} \Phi^{U}}{E_j S_b f_N} + \\
&\frac{R_{u_j,t} (
T_{\text{QSS}} 
-\Phi_{db}^{U}
-\frac{1}{2} \Phi^{U})}{E_j}, \quad
\forall{j \in \mathcal{N}_{dc}}, \forall{t \in \mathcal{T}}
\end{aligned}
\label{eq:ed_model15}
\end{equation}
\begin{equation}
\begin{aligned}
&\underline{e}_j \leq 
e_{j,t+1} - \frac{M_{u_j,t} S_{u_j} \cdot \mathrm{RoCoF}_{\lim} \Phi^{U}}{E_j S_b f_N} -  \\
&\frac{R_{u_j,t} (
T_{\text{QSS}} 
-\Phi_{db}^{U}
-\frac{1}{2} \Phi^{U})}{E_j} \leq \overline{e}_j
, \quad
\forall{j \in \mathcal{N}_{dc}}, \forall{t \in \mathcal{T}}
\end{aligned}
\label{eq:ed_model16}
\end{equation}
where the decision sets are defined as
$$
\begin{aligned}
\Xi_g &= \left\{p_{g_i,t}, R_{g_i,t} \mid i \in \mathcal{N}_g,\; t \in \mathcal{T}\right\}, \\
\Xi_{dc} &= \left\{p_{u_j,t}, R_{u_j,t}, M_{u_j,t}, e_{j,t+1} \mid j \in \mathcal{N}_{dc},\; t \in \mathcal{T}\right\}, \\ 
\Xi_{agg} &= \left\{R_{G,t}, R_{U,t} \mid t \in \mathcal{T}\right\}. 
\end{aligned}
$$
Moreover, $a_{g_i}$ and $a_{u_j}$ denote the generation cost coefficient of SG $i$ and the charging/discharging degradation cost coefficient of UPS $j$, respectively; $b_{g_i}$ and $b_{u_j}$ denote the corresponding PFR service cost coefficients; and $c_{u_j}$ denotes the virtual inertia service cost coefficient of UPS $j$. Parameters $\overline{p}_{g_i}$, $\underline{p}_{g_i}$, and $\overline{R}_{g_i}$ are the maximum generation limit, minimum generation limit, and maximum available PFR reserve capacity of generator $i$, respectively. Similarly, $\overline{p}_{u_j}$, $\underline{p}_{u_j}$, and $\overline{M}_{u_j}$ denote the maximum discharging power limit, maximum charging power limit, and maximum virtual inertia parameter limit of UPS $j$, respectively. Parameter $E_j$ is the energy capacity of UPS $j$; $\underline{e}_j$ and $\overline{e}_j$ denote the safe minimum SoC limit reserved for emergency backup operation and the maximum SoC limit, respectively. $\underline{R}_{G,t}$ denotes the minimum aggregate PFR reserve capacity requirement provided by SGs. 
And 
$\Delta t$ denotes the duration of each scheduling interval
.

The objective function in Eq.~(\ref{eq:ed_model1}) minimizes the total operation cost, including the generation cost and PFR service cost of SGs, as well as the charging/discharging degradation cost, PFR service cost, and inertia service cost of DC UPS systems. Constraint (\ref{eq:ed_model2}) enforces system power balance with UPS participation in steady-state power balancing. Constraints (\ref{eq:ed_model3})--(\ref{eq:ed_model6}) describe SG operating and PFR requirements. 
Constraint (\ref{eq:ed_model7}) guarantees that the frequency nadir occurs during the SG PFR ramping stage, where both SGs and UPS systems contribute to frequency support.
Constraint (\ref{eq:ed_model8}) limits UPS charging/discharging power, while Constraint (\ref{eq:ed_model9}) allocates UPS power capacity among steady-state power support, PFR service, and virtual inertia response. Constraints (\ref{eq:ed_model10})--(\ref{eq:ed_model11}) enforce UPS PFR and virtual inertia limits. Finally, Constraints (\ref{eq:ed_model12})–(\ref{eq:ed_model16}) model the UPS SoC evolution, enforce the SoC operating limits and cyclic SoC condition, guarantee sufficient energy availability for the scheduled inertia and PFR services, and ensure that the post-response SoC remains within the prescribed safety range.
The proposed optimization model is formulated as a SOCP problem and can be efficiently solved by existing off-the-shelf convex optimization solvers.

\section{Numerical Validation}
\subsection{System Description}
\begin{figure}[t]
\centering
\includegraphics[width=3.4in]{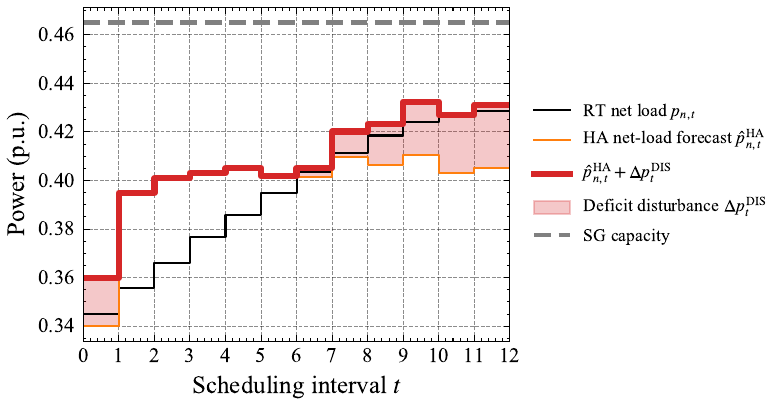}
\caption{One-hour net-load and deficit disturbance profiles.}
\label{fig:data}
\end{figure}

To evaluate the effectiveness of the proposed RTED model, a modified IEEE 39-bus system \cite{she2023virtual} consisting of six SGs and four DC UPS systems is considered. In the simulation, each DC UPS system can be interpreted as a UPS aggregator, i.e., a cluster of multiple UPS units under centralized coordination. Therefore, the rated capacity of each DC UPS system reflects the aggregate power capability of multiple UPS units rather than that of a single UPS unit.
The system power base is selected as $10^4$ MVA. 
The generation cost coefficients $a_{g_i}$ and $b_{g_i}$ are set to $[19,20,21,22,24,17]$ and $[8.2,8.4,8.6,8.8,9.2,7.5]\times10^4/(\text{p.u.}\cdot\text{h})$, respectively. 
The UPS degradation cost coefficients $a_{u_j}$, PFR reserve cost coefficients $b_{u_j}$, and inertia-response cost coefficients $c_{u_j}$ are specified as $[2.0,2.2,2.4,2.6]\times10^4\$/(\text{p.u.}^2\cdot\text{h})$, $[3.0,3.2,3.4,3.6]\times10^4\$/(\text{p.u.}\cdot\text{h})$, and $[0.02,0.025,0.03,0.035]\times10^4\$/(\text{s}\cdot\text{h})$, respectively. 
The frequency-security requirements are characterized by a maximum RoCoF limit of 0.5 Hz/s and a frequency nadir limit of 0.4 Hz. A one-hour scheduling horizon with twelve 5-minute intervals is considered. The proposed optimization model is implemented in CVXPY and solved using a convex optimization solver. 
As illustrated in Fig.~\ref{fig:data}, the hour-ahead net-load forecast profile is constructed based on scaled CAISO operational data. The active-power deficit disturbance is modeled as the 95\%-confidence worst-case error caused by renewable generation over-forecast.

\subsection{Scheduling Results}
\begin{figure}[t]
\centering
\includegraphics[width=3.55in]{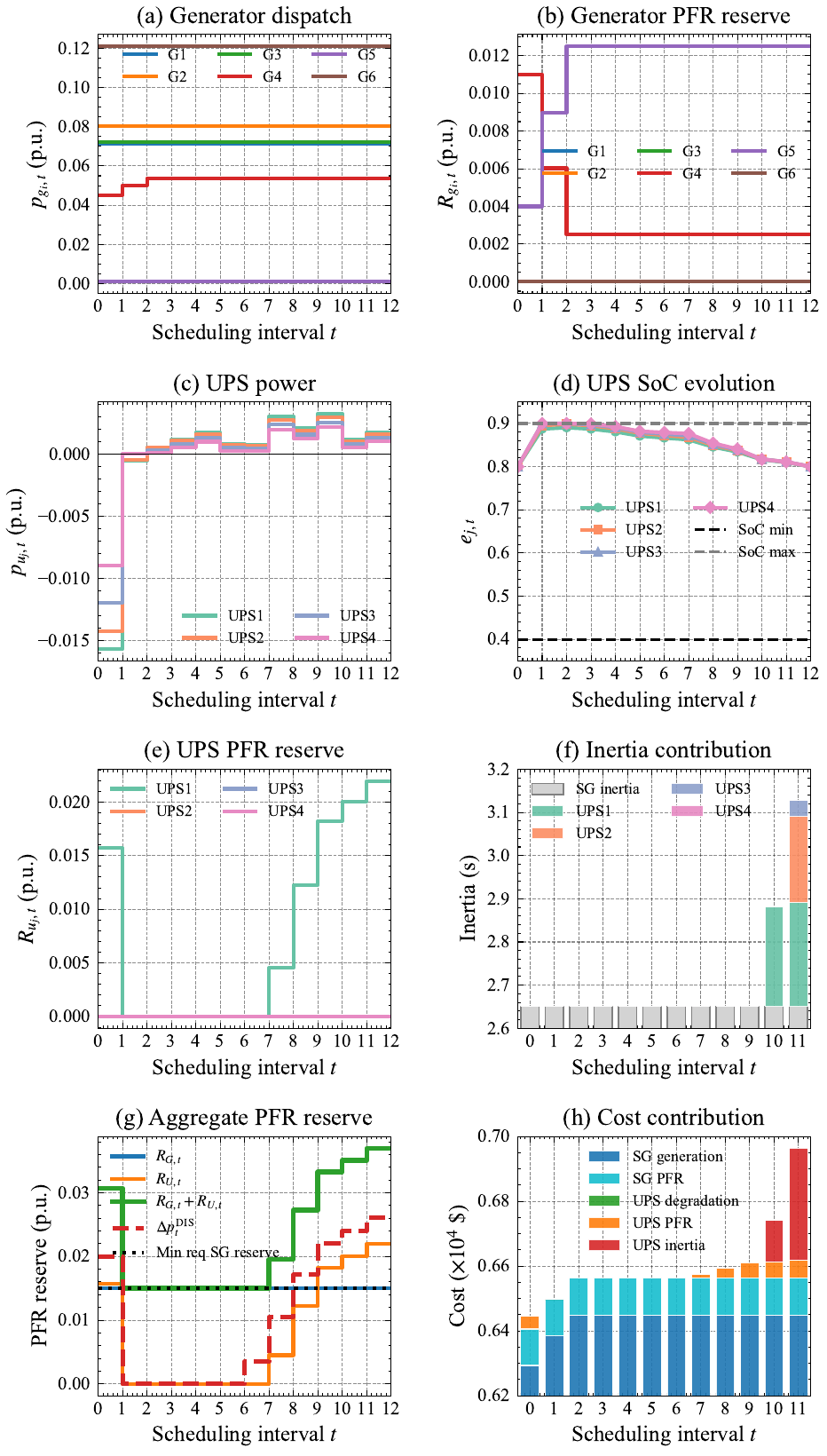}
\caption{Illustration of the scheduling results}
\label{fig:results}
\end{figure}

The total scheduling cost over the one-hour RTED horizon is \$79,245. As shown in Fig.~\ref{fig:results}, lower-cost SGs are preferentially dispatched to satisfy most of the system demand, while the remaining generation capacity is utilized to provide PFR reserves according to reserve costs. Since battery degradation costs are significantly lower than generation costs, the UPS systems are charged during low-load periods and discharged during high-load periods to reduce generation costs, resulting in a load-leveling effect. The UPS SoC returns to its initial value at the end of the scheduling horizon. 
Since the reserve cost of UPS systems is lower than that of SGs, the model maintains only the minimum required SG reserve and does not schedule additional SG reserves beyond this requirement.
Owing to its lowest reserve cost, UPS1 exclusively provides additional PFR reserves whenever the minimum SG reserve requirement becomes insufficient. During most intervals, the inertia provided by SGs alone is adequate. However, under severe disturbances at $t=10$ and $t=11$, additional virtual inertia support from UPS systems is economically activated. Moreover, the total PFR reserve requirement exceeds the disturbance magnitude because part of the reserve is used to compensate for the delayed response of conventional generators. Finally, generation cost dominates the total scheduling cost, whereas the costs associated with UPS degradation and PFR support remain relatively small. The UPS inertia response cost becomes noticeable only during intervals with large disturbances, reflecting the premium associated with fast inertial support.
Note that although both services are provided by UPS systems, inertia response  requires maintaining sufficient headroom and allocating additional converter capability for ultrafast response, making its marginal cost higher than that of PFR support.

\subsection{Frequency Security Analysis}
Table.~\ref{tab:freq_index} compares the frequency-security performance with and without UPS support. For a fair comparison, the same SG PFR reserve schedule obtained from the proposed RTED is adopted in both cases. In the w/o UPS case, only SGs provide inertia and PFR support, while all UPS inertia and PFR contributions are removed. As shown in Table.~\ref{tab:freq_index}, the proposed UPS-assisted frequency support reduces the maximum RoCoF from 0.5900 Hz/s to the prescribed limit of 0.5000 Hz/s, corresponding to a 15.26\% improvement. More importantly, the maximum frequency deviation is reduced from 5.4216 Hz to 0.4000 Hz, yielding a 92.62\% reduction and effectively preventing UFLS. These results demonstrate the effectiveness of DC UPS systems in enhancing frequency security under active-power deficit disturbances.
\begin{table}[!t]
\caption{Frequency-security performance comparison.}
\label{tab:freq_index}
\centering
\small
\setlength{\tabcolsep}{3pt}
\renewcommand{\arraystretch}{1.2}
\begin{tabular}{lcccc}
\hline
Metric &
w/o UPS &
w/ UPS &
\makecell{Abs.\\Gain} &
\makecell{Rel.\\Gain (\%)}\\
\hline
\makecell[l]{Maximum\\RoCoF (Hz/s)}
&
0.5900
&
0.5000
&
0.0900
&
15.26
\\
\makecell[l]{Maximum frequency\\deviation (Hz)}
&
5.4216
&
0.4000
&
5.0216
&
92.62
\\
\hline
\end{tabular}
\end{table}

\section{Conclusion}
This paper proposed a frequency-security-aware RTED model that coordinates SGs and DC UPS systems for energy balancing, PFR, and inertia response services. Analytical frequency-security constraints were derived and incorporated into the dispatch framework through SOCP formulation. Numerical results demonstrated that the proposed model with UPS flexibility can economically provide fast frequency support while reducing generation costs through UPS capacity allocation and energy shifting. Moreover, the proposed model reduced the maximum RoCoF by 15.26\% and the maximum frequency deviation by 92.62\%, effectively preventing UFLS under active-power deficit disturbances. 

Future work will extend the proposed RTED model to more general disturbance scenarios, including both active-power deficit and surplus disturbances, and investigate the optimal settings of the initial SoC and minimum SoC limits for DC UPS systems. Moreover, suitable incentive mechanisms need to be designed to encourage the voluntary participation of UPS systems in frequency support ancillary service markets.

\IEEEtriggeratref{1}
\IEEEtriggercmd{\enlargethispage{2\baselineskip}}
\bibliographystyle{IEEEtran}
\bibliography{reference}

\end{document}